\documentstyle[12pt]{article}
\font\frbig=eufb10  scaled\magstep1
\font\frscr=eufb7 scaled\magstep1
\font\frscrscr=eufb5 scaled\magstep1

\newfam\frfam
\textfont\frfam=\frbig
\scriptfont\frfam=\frscr
\scriptscriptfont\frfam=\frscrscr
\def\fr{\fam\frfam}

\font\openbig=msym10  scaled\magstep1
\font\openscr=msym7 scaled\magstep1
\font\openscrscr=msym5 scaled\magstep1

\newfam\openfam
\textfont\openfam=\openbig
\scriptfont\openfam=\openscr
\scriptscriptfont\openfam=\openscrscr
\def\open{\fam\openfam}

\font\Scbig=cmss10  scaled\magstep1
\font\Scscr=cmss8 scaled\magstep1
\font\Scscrscr=cmss8

\newfam\Scfam
\textfont\Scfam=\Scbig
\scriptfont\Scfam=\Scscr
\scriptscriptfont\Scfam=\Scscrscr
\def\Sc{\fam\Scfam}

\makeatletter

\newdimen\normalarrayskip              % skip between lines
\newdimen\minarrayskip            % minimal skip between lines
\normalarrayskip\baselineskip
\minarrayskip\jot
\newif\ifold             \oldtrue     \def\new{\oldfalse}
\def\arraymode{\ifold\relax\else\displaystyle\fi}
\def\@arrayskip{\ifold\baselineskip\z@\lineskip\z@
     \else
     \baselineskip\minarrayskip\lineskip2\minarrayskip\fi}
\def\@arrayclassz{\ifcase \@lastchclass \@acolampacol \or
\@ampacol \or \or \or \@addamp \or
   \@acolampacol \or \@firstampfalse \@acol \fi
\edef\@preamble{\@preamble
  \ifcase \@chnum
     \hfil$\relax\arraymode\@sharp$\hfil
     \or $\relax\arraymode\@sharp$\hfil
     \or \hfil$\relax\arraymode\@sharp$\fi}}
\def\@array[#1]#2{\setbox\@arstrutbox=\hbox{\vrule
     height\arraystretch \ht\strutbox
     depth\arraystretch \dp\strutbox
     width\z@}\@mkpream{#2}\edef\@preamble
{\halign \noexpand\@halignto
\bgroup \tabskip\z@ \@arstrut \@preamble \tabskip\z@ \cr}%
\let\@startpbox\@@startpbox \let\@endpbox\@@endpbox
  \if #1t\vtop \else \if#1b\vbox \else \vcenter \fi\fi
  \bgroup \let\par\relax
  \let\@sharp##\let\protect\relax
  \@arrayskip\@preamble}

\makeatother

\begin{document}
\hfuzz=1pt

\date{(November 25, 1991)}
\title{{\sc Kontsevich - Miwa Transform of the
Virasoro Constraints\\
as Null-Vector Decoupling Equations}
\footnote{This is a revised
version of the author's virtual paper {\sc A
Kontsevich - Miwa Transform of the
Virasoro Constraints on KP and Generalized
K{\rm d}V Hierarchies} (Oct.
1991) which consisted to a considerable degree
of arithmetical errors and is herewith nullified.}}
        \author{{\large A.M.Semikhatov}\\
\sl Theory Division, P.N.Lebedev Physics Institute\\
\sl 53 Leninsky prosp., Moscow SU 117924 USSR}

\maketitle

\begin{abstract}
We use the Kontsevich--Miwa transform to relate the
Virasoro constraints on integrable hierarchies
with the David-Distler-Kawai formalism of
gravity-coupled
conformal theories. The derivation
relies on evaluating the energy-momentum tensor
on the hierarchy at special
values of the spectral parameter.
We thus obtain in the Kontsevich parametrization the
`master equations' which implement the Virasoro
constraints and at the
same time coincide with null-vector decoupling
equations in an `auxiliary'
conformal field theory on the complex plane of
the spectral parameter.
This gives the operators their gravitational scaling
dimensions
(for one out of four possibilities to choose signs),
with the
$\alpha_+$ being equal to the background charge
$Q$ of an abstract $bc$ system underlying the
structure of the Virasoro constraints.
The formalism also generalizes to the
$N$-KdV hierarchies.
\end{abstract}

\newpage
 \section{Introduction}

\leavevmode\hbox to\parindent{\hfill}
The interest in Matrix Models \cite{[BK],[DSh],[GM]}
has been stimulated,
besides their
applications to matter+gravity systems \cite{[DVV],
[GGPZj],[DfK],[FKN2]}, also
by intriguing relations they have with completely
integrable equations and the
intersection theory on the moduli space of curves
\cite{[D],[W1],[P],[K],[W2]}.
On the other hand, as to the foundation of the matrix model
approach by itself, a challenging problem is the direct
proof of its
equivalence
to the conformal field theory formalism for
quantum gravity \cite{[KPZ],[DK],[Da]}. Assuming that this
equivalence exists,
as is suggested by a circumstantial evidence,
one then has to believe that certain ingredients of conformal
field theory
satisfy integrable equations.
These, however, seem to be a long way from the equations
which are known to hold
for conformal field theory correlators \cite{[DF],[KZ]}.

As we will show, the role of an `intermediary'
on the way from non-linear KdV-like equations
to conformal field theory is played by the Virasoro
constraints on integrable hierarchies (which in another
guise are recursion
relations in topological theories) \cite{[FKN1],[DVV],
[IM],[MM]}. The
Virasoro
constraints are the heart of the matrix models' applications
to both
gravity-coupled theories and the intersection theory.
The case studied in most detail is the
Virasoro-constrained KdV hierarchy whose relation to the
intersection theory on
moduli space of Riemann surfaces has been discussed
in \cite{[W1]}.

To reveal this role of the Virasoro constraints, we
will adopt the approah which has proved fruitful in
establishing the relation
between matrix models and the
universal moduli space: That is, we borrow from the
matrix model due to
Kontsevich  the choice of independent variables. The
Kontsevich matrix model
provides a combinatorial model of the universal moduli
space
\cite{[K]} and, as such, serves as an important step in
demonstrating the KdV
hierarchy in the intersection theory on the moduli space.
Note, however,
that the
Kontsevich model is not of the form of the matrix models
considered previously,
which raises the question of its equivalence to one of the
``standard" models.
The crucial point in studying this equivalence is,
again, the proof of the Virasoro
constraints satisfied by the Kontsevich matrix integral
\cite{[W2],[MMM]}. Once
the constraints are established, one is left with ``only"
the proof that they
specify the model uniquely.

The Kontsevich model by itself, as well as the existing
derivations of the
Virasoro constraints,
appear to be tied up to the KdV hierarchy and thus to
the $l = 2$
minimal models. There exist, at the same time, matrix
models of other minimal
conformal theories coupled to gravity, which correspond
to higher generalized
`KdV' hierarchies \cite{[DS]}
\footnote{ In this paper we will restrict ourselves to
the series associated to
$sl(N)$ Kac-Moody algebras (and therefore the {\it A}-series
minimal models \cite{[CIZ]}, and we will call these the
$N$-KdV hierarchies.
Virasoro constraints on the $N$-KdV hierarchies admit a unified
treatment, which is in turn a specialization of a general
construction
applicable to hierarchies of the $r-$matrix type \cite{[S6]}.}.
 Although neither
 the interpretation of Virasoro-constrained
$N$-KdV hierarchies in terms of moduli spaces, nor the
corresponding Kontsevich-type matrix integrals are known,
it would not be
natural if the $N=2$ case were exceptional. Thus
another problem is
whether a unified approach exists which allows one
to recast Virasoro
constraints on $N$-KdV (and hopefully other) hierarchies
into Ward
identities of a would-be Kontsevich-type matrix integral.

In this paper we take as a starting point the Virasoro
constraints in the usual
parametrization and then investigate
whether they can be recast into the Kontsevich variables.
There are
obvious similarities between Kontsevich' parametrization
and the Miwa transform
used in the KP hierarchy \cite{[Sa]}. We thus attempt to proceed
with a general
Miwa transformation. As we will see, the Virasoro constraints
are {\it not}
reformulated nicely unless one restricts the Miwa transformation
to the
Kontsevich one, yet we find it instructive to see at which step
the Miwa
parametrization fails to work.

Pulling back the Virasoro generators to the Kontsevich
parametrization seems only possible for the combination
$\sum _{n\geq -1} L_n z^{-n-2}$ of the Virasoro generators,
and only at special values of the spectral parameter $z$.
The resulting relations are candidates for the Ward
identities corresponding to a Kontsevich-type matrix
integral.  For the $N$-KdV hierarchies these relations are
the analogues of the ``master equation'' of ref.\cite{[MS]}.
Further, we show in the KP case that our version of the
master equation (which, though very similar to, is not quite
the same as the one from \cite{[MS]}) happens to be nothing
but an  equation on correlation functions in an `auxiliary'
conformal field theory, stating the decopling of a certain
null vector. This is in the classical spirit of
\cite{[BPZ]}, yet the applicability to Virasoro-constrained
hierarchies seems to be new. Recall in particular that
according to the basic matrix models ideology, the resulting
equations that hold on an abstract ({\it
spectral-parameter}) ${\open CP}^1$ are in fact
non-perturbative.  It is thus very reassuring to recover in
our approach the results of refs.\cite{[DK],[Da]}!

\section{Virasoro action on the KP hierarchy}

 \leavevmode\hbox to\parindent{\hfill}{\bf 2.1}. The KP
hierarchy is described in terms of $\psi {\rm Diff}$
operators \cite{[DDKM]} as an infinite set of mutually
commuting evolution equations $$ {\partial K\over \partial
t_r} = - (KD^rK^{-1})_-K,\qquad  r \geq  1 \eqno{(2.1)} $$
on the coefficients $w_n(x${\bf ,} $t_1$, $t_2$,
$t_3,\ldots)$ of a $\psi {\rm Diff}$ operator (more
precisely, a $\psi {\rm Diff}$ symbol) $K$ of the form (here
and in the sequel, $D = \partial /\partial x )$ $$ K = 1 +
\sum _{n\geq 1}w_n D^{-n} \eqno{(2.2)} $$

Introduce a `matrix model potential' $$
\xi (t,z) = \sum _{r\geq 1}t_r z^r \eqno{(2.3)}$$ The wave
function and the adjoint wave function are then defined by
$$ \psi (t,z) = Ke^{\xi (t,z)} ,\qquad  \psi ^\ast (t,z) =
K^{\ast -1}e^{-\xi (t,z)} \eqno{(2.4)} $$ where $K^\ast $ is
the formal adjoint to $K$. The wave function $\psi $ is an
eigenfunction of the Lax operator:  $$ Q\psi (t,z) = z\psi
(t,z),\qquad         Q \equiv  KDK^{-1} \eqno{(2.5)}$$
functions. The basic property of the wave functions is their
relation to the tau function:  $$ \psi(t,z) = e^{\xi
(t,z)}{\tau (t-[z^{-1}])\over \tau (t)},\quad \psi^\ast
(t,z) = e^{-\xi (t,z)}{\tau (t+[z^{-1}])\over \tau (t)}
\eqno{(2.6),(2.7)} $$ where, $$ t \pm  [z^{-1}] = (t_1 \pm
z^{-1}, t_2 \pm  {1\over 2} z^{-2}, t_3 \pm  {1 \over 3}
z^{-3}, \ldots) \eqno{(2.8)}$$

{\bf 2.2.} Now we introduce a Virasoro action on the tau
function: The Virasoro generators read, $$ \new
\begin{array}{rcl} {\Sc L}_{p>0} &=& {1\over 2}\sum
^{p-1}_{k=1}{\partial^{2} \over \partial t_{p-k} \partial
t_k} + \sum _{k\geq 1}kt_k {\partial \over \partial t_{p+k}}
+ (a_0 + (J - {1\over 2})p) {\partial \over \partial t_p}\\
{\Sc L}_0&=& \sum _{k\geq 1}kt_k {\partial \over \partial
t_k} + {1\over 2}a^2_0 - {1\over 24} \\ {\Sc L}_{p<0} &=&
\sum _{k\geq 1}(k - p)t_{k-p} {\partial \over \partial t_k}+
{1\over 2}\sum ^{-p-1}_{k=1}k(-p - k)t_kt_{-p-k}+ (a_0 + (J
- {1\over 2})p)(-p)t_{-p} \end{array} \eqno{(2.9)} $$ They
satisfy the algebra $$ [{\Sc L}_p, {\Sc L}_q ] = (p - q)
{\Sc L}_{p+q} + \delta _{p+q,0}(-p^3)(J^2 - J + {1\over 6})
\eqno{(2.10)}$$ which shows, in particular, the role played
by the parameter {\it J}.  (Shifting $ {\Sc L}_0$ as $ {\Sc
L}_0 \mapsto   {\Sc L}_0 - {1\over 2} (J^2 - J + {1\over
6})$ we recover in (2.10) the standard central term $-\delta
_{p+q,0}(p^3-p)(J^2 - J + {1\over 6})$.) It will be quite
useful to introduce an ``energy-momentum tensor" $$ {\Sc T}
 (u) = \sum _{p\in  { \open Z}}u^{-p-2} {\Sc L}_p
 \eqno{(2.11)} $$ Using this to deform the tau function as
$$ \tau (t) \mapsto \tau (t) + \delta \tau (t) = \tau (t) +
{\Sc T} (u)\tau (t), \eqno{(2.12)} $$ we translate this
action into the space of dressing operators {\it K}. The
result is \cite{[S1]} that $K$ gets deformed by means of a
left multiplication, $$ \delta K = - {\fr T}(u)K, $$ where
${\fr T}(u)$ is the energy-momentum tensor in another guise,
now a pseudodifferential operator \footnote{We have chosen
the irrelevant parameter $a_0 = {1\over 2}$, see \cite{[S1]}
where $a_0 = N+{1\over 2}$.} $$ {\fr T}(u) = (1 - J)
 {\partial \psi (t,u)\over \partial u} \circ  D^{-1} \circ
\psi ^\ast (t,u) - J\psi (t,u) \circ  D^{-1} \circ {\partial
\psi ^\ast (t,u)\over \partial u} \eqno{(2.13)} $$ Thus, $
{\fr T}(u)$ reproduces the structure of the energy-momentum
tensor of a spin$-J\ bc$ theory \cite{[FMS]}. In its own
turn, $ {\fr T}(u)$ can be expanded in powers of the
variable $u$, which was introduced in (2.11) and has now
 acquired the role of a spectral parameter, as $$ {\fr  T}
(u) = \sum _{p\in {\open Z}}u^{-p-2} {\fr L}_p \eqno{(2.14)}
$$ This gives the individual Virasoro generators (which are
a particular case of the general construction applicable to
integrable hierarchies of the $r-$matrix type \cite{[S6]})
$$ {\fr L}_r \equiv K(J(r+1)D^r + PD^{r+1})K^{-1})_-,\quad
 P \equiv x + \sum _{r\geq 1}rt_r D^{r-1} \eqno{(2.15)} $$

{\bf 2.3.} The KP hierarchy can be reduced to generalized
$N$-KdV hierarchies \cite{[DDKM]} by imposing the constraint
$$ Q^N \equiv  L \in  {\rm Diff} \hspace{3em}  (
\Rightarrow  Q^{Nk} \in  {\rm Diff}\hbox{, } k \geq  1)
\eqno{(2.16)}$$ requiring that the $N^{th}$ power of the Lax
operator be purely differential.  Then, in a standard
manner, the evolutions along the times $t_{Nk}$ , $k \geq
1$, drop out and these times may be set to zero. The rest of
the $t_n$ are conveniently relabelled as $t_{a,i} =
t_{Ni+a}$,  $i \geq  0$,  $a = 1,\ldots,N-1.$

As to the Virasoro generators, only $ {\fr L}_{Nj}$ are
compatible with the reduction in the sense that they remain
symmetries of the reduced hierarchy without imposing any
further constraints \cite{[S3]}. The value of $J$ can be set
to zero \cite{[S3],[S4]}. Then, after a rescaling, the
 generators $$ {\fr L}^{[N]}_j = {1\over N} (K\sum _{a,i}(Ni
+ a) t_{a,i} D^{N(i+j)+a} K^{-1})_- \eqno{(2.17)} $$ span a
Virasoro algebra of their own.

Again, we find it very useful to construct the
energy-momentum tensor corresponding to these generators.
Recall that the spectral parameter of the $N$-KdV hierarchy
is  $\zeta  = z^N$. Then $$ \new \begin{array}{rcl} {\fr
T}^{[N]}(\zeta )(d\zeta )^2 &\equiv& \sum_{j\in  {\open
 Z}}\zeta ^{-j-2}  {\fr L}^{[N]}_j(d\zeta )^2\\ {}&=&
N\left( K\sum_{b,j}(Nj+b) t_{b,j} D^{Nj+b} {1\over z^2}
\delta (D^N,z^N) K^{-1}\right) _-(dz)^2 \end{array}
\eqno{(2.18)} $$ where $$ \delta (u,v) = \sum_{n \in  {\open
Z}}\left( {u\over v} \right) ^n \eqno{(2.19)} $$ denotes the
formal delta function. The essential point is that $\delta
(z,D)$ is a projector onto an eigenspace of $D$ with the
eigenvalue $z$. Then it is obvious that $$ \delta (D^N,z^N)
= {1\over N} \sum ^{N-1}_{c=0}\delta (z^{(c)}, D),\quad
z^{(c)} = \omega ^cz,\quad  \omega  = \exp \left( {2\pi
\sqrt{-}1\over N}\right) \eqno{(2.20)} $$ Using this we
bring the above energy-momentum tensor to the form $$ {\fr
 T}^{[N]}(E) =\sum ^{N-1}_{c=0}\omega ^c {\partial \psi
(t,z^{(c)})\over \partial z} \circ  D^{-1} \circ \psi ^\ast
(t,z^{(c)}) = {1\over N}\sum ^{N-1}_{c=0}\omega ^{2c}  {\fr
T}(z^{(c)}) \eqno{(2.21)} $$ where the following notations
have been used: recall that the spectral parameter of an
$N$-KdV hierarchy lies on a complex curve defined in $  {
\open C}^2 \ni (z,E)$ by an equation $z^N = P(E)$. Then,
$\psi $ and $\psi ^\ast $ are defined on this curve, and
after the projection onto $  {\open CP}^1$ yield $N$ wave
functions $\psi ^{(a)}(t,E)$, distinct away from the branch
points. That is, we have defined $$ \psi ^{(a)}(t,E) =
Ke^{\xi (t,z^{(a)})} \equiv w(t,z^{(a)})e^{\xi
(t,z^{(a)})},\qquad \xi (t, z^{(a)}) = \sum _{j,b}t_{b,j}
(z^{(a)})^{Nj+b} \eqno{(2.22)} $$ Note a striking similarity
between (2.21) and the energy-momentum tensor of conformal
theories on $ { \open Z}_N-$curves \cite{[BR]}.

{\bf 2.4.} To conclude this review, we outline the basic
steps of how the Virasoro action on the $N-$KdV tau
function, by the generators of the type of (2.9), can be
recovered from the ``energy-momentum tensor" (2.21).  The
usual way to derive objects pertaining to the tau function
is through the use of the equation

$$ {\rm res} K = - \partial \log \tau ,
$$ whence $$ \delta \partial \log \tau  = - {\rm res} \delta
K = {\rm res}  {\fr T}^{[N]}(z)K = {\rm res} {\fr
T}^{[N]}(z) \eqno{(2.23)} $$ The residue of $ {\fr
T}^{[N]}(z)$ is immediately read off from (2.21). To the
combination of wave functions thus appearing we apply the
formula $$ \partial \left( {1\over u-z}e^{\xi (t,u) - \xi
(t,z)} {\tau (t + [z^{-1}] - [u^{-1}])\over \tau (t)}
\right) = \psi (t,u)\psi ^\ast (t,z) \eqno{(2.24)} $$ (it
follows directly by applying the vertex operator $\exp
\sum_{r\geq 1}{1\over r} (z^{-r}-u^{-r}) {\partial \over
\partial t_r}$ to the bilinear identity of the KP hierarchy
\cite{[DDKM]} and then evaluating the integral as a sum over
residues). It follows from (2.24) by expanding it at $z
\rightarrow  u$ that $$ \new \begin{array}{rcl} {\partial
\psi (t,z)\over \partial z} \psi ^\ast (t,z) &=& \partial
\left\{ {1\over 2} {1\over \tau (t)} {1\over z} \nabla (t,z)
 {1\over z} \nabla (t,z)\tau (t) + {1\over \tau (t)}
{\partial \xi (t,z)\over \partial z} {1\over z} \nabla
 (t,z)\tau (t) \right. \\ &+& \left.  {1\over 2} {1\over
\tau (t)} {1\over z} \nabla  (t,z)\tau (t) + {1\over
2}\left( {\partial \xi (t,z)\over \partial z}\right) ^2
\right\} \end{array} \eqno{(2.25)} $$

This expression by itself would lead us back to the KP
Virasoro generators (2.9) (with $J=0)$. Now the $N -$KdV
generators follow according to the formula (2.21), by
substituting $z \mapsto  z^{(c)}$ and summing over $c$. The
sum over $c$ plays the role of a projector onto the identity
of the group of $N^{th}$ roots of unity. Therefore, $$ \new
\begin{array}{l} \sum ^{N-1}_{c=0}\omega ^{2c} {\partial
\psi (t,z^{(c)})\over \partial z^{(c)}} \psi ^\ast
 (t,z^{(c)}) = N \partial \left\{{1\over 2}\sum ^{N-1}_{a=1}
\sum _{i,j\geq 0}(Nj + a)(N(j+1) - a)t_{a,i} t_{N-a,j}
z^{N(i+j+1)-2}\right. \\ \hspace{1em} \left. + {1\over 2}
{1\over \tau (t)}\sum ^{N-1}_{a=1}\sum_{i,j\geq
0}z^{-N(i+j+1)-2} {\partial^2\tau (t)\over \partial
t_{a,i}\partial t_{N-a,j}} + {1\over \tau (t)} \sum
 ^{N-1}_{a=1}\sum _{i,j\geq 0}(Ni + a)t_{a,i} z^{N(i-j)-2}
{\partial \tau (t)\over \partial t_{a,j}} \right\}
\end{array} \eqno{(2.26)} $$ from which the Virasoro
generators can be read off as $$ \new \begin{array}{rcl} n >
0:  \qquad   {\Sc L} ^{[N]}_n &=& {1\over N} {1\over 2}\sum
^{N-1}_{a=1}\sum ^{n-1}_{i=0}{\partial ^2\over \partial
t_{a,i} \partial t_{N-a,n-i-1}} + {1\over N}\sum
^{N-1}_{a=1}\sum _{i\geq 0}(Ni + a)t_{a,i} {\partial \over
\partial t_{a,i+n}},\\ {\Sc L} ^{[N]}_0 &=& {1\over N}\sum
 ^{N-1}_{a=1}\sum _{i\geq 0}(Ni + a)t_{a,i} {\partial \over
\partial t_{a,i}},\\ n < 0:  \qquad  {\Sc L} ^{[N]}_n &=&
{1\over N} {1\over 2}\sum^{N-1
}_{a=1}\sum^{-n-1}_{i=0}(Ni+a)(-N(i+n)-a) t_{a,i}
t_{N-a,-i-n-1}\\ {}&+&{1\over N}\sum ^{N-1}_{a=1}\sum
_{i\geq -n} (Ni+a)t_{a,i} {\partial \over \partial
t_{a,i+n}} \end{array} \eqno{(2.27)} $$ These generators act
on the tau functions $\tau (t)$ of the $N$-KdV hierarchy.

\section{Miwa--Kontsevich transform} \leavevmode\hbox
to\parindent{\hfill}Now we are going to use the same
strategy as was used to derive (2.27), but this time in the
Miwa--Kontsevich parametrization of the times of the
hierarchy. As in the above, we start with the simplest case,
the KP hierarchy.

The Miwa reparametrization of the KP times is accomplished
by the substitution $$ t_r = {1\over r} \sum  _j n_j
z^{-r}_j \eqno{(3.1)}$$ where $\{z_j \}$ is a set of points
on the complex plane and $n_j$ are integers. This
parametrization puts, in a sense, the times and the spectral
parameter on equal ground. It may in some cases be
conceptually advantageous to write (3.1) as $$ t_r = {1\over
r} \sum _{ z \in {\open CP}^1} n_z z^{-r} \eqno{(3.2)} $$
where $n_z$ is nonvanishing for only a finite (countable)
set of points. Then the tau function becomes a functional
$\tau [n]$ of a function $n$ on $  {\open C}  {\open P}^1$
which must be from the class of functions in some sense
close to linear combinations (with integer coefficients) of
delta-functions.  On the other hand, the way Kontsevich has
used a parametrization of this type implied setting all the
$n_j$ equal to unity. We will in fact see why a restriction
of this kind is necessary, but this will require working
with the general $n_j$ for as long as possible.

The Miwa substitution turns out very inconvenient with
regard to the use of the standard machinery of the KP
hierarchy $(e.g.$, proceeding along the usual chain (tau
function$) \mapsto  ($wave function$) \mapsto  ($dressing
operator), etc.); instead, it serves to construct a quite
different, ``discrete'' formalism for the KP and related
hierarchies \cite{[Sa]}. Now, the above expressions (2.15),
(2.17) and (2.21) for the Virasoro generators involve all
the standard ingredients such as the wave functions, the
spectral parameter etc., which complicates re-expressing
them in the Miwa parametrization. That is, taking (3.2)
seriously, and even viewing it as $$ t_r = {1\over r} \int
_{ {\open CP}^1} d\mu (z) z^{-r} n(z) \eqno{(3.3)} $$ one
can formally define the wave functions as $$ \psi [n](z) =
\prod_j \left(1-{z\over z_j} \right)^{n_j} {1\over \tau [n]}
e^{{\delta \over \delta n(z)}} \tau [n] \eqno{(3.4)} $$
Short-`distance' expansion as in (2.24) -- (2.25) would then
require making sense out of this formula and similarly out
of expressions such as ${\partial \over \partial z} {\delta
\over \delta n[z]}$. Even this, however, would not be quite
satisfactory, as one would still have had to express the
result in terms of the derivatives with respect to $z_j$ ,
which are the parameters of the Kontsevich model: for us,
the tau function must be a function $\tau \{z_j \}$ of
points scattered over $  {\open CP}^1 $.

There are two circumstances, however, that save the day.
First, we will be interested not in all the Virasoro
generators, but rather in those with non-negative (and, in
addition, $-1$) mode numbers $ {\fr L}_{n\geq -1}$
\footnote{\ It is these Virasoro generators that are used to
 define {\it Virasoro-constrained} hierarchies, simply as $
{\fr L}_n = 0$, $n \geq -1$.}.  Picking these out amounts to
retaining in $ {\fr T}(z)$ only terms with $z$ to negative
 powers, $i.e.$, the terms vanishing at $z \rightarrow
\infty $. This part of $ {\fr T}(z)$ is singled out as $$
{\fr T}^{(\infty )}(v) = \sum _{n\geq 0}v^{-n-1} {1\over
2\pi i} \oint dz z^n  {\fr T}(z) = {1\over 2\pi i} \oint dz
{1\over v - z}  {\fr T}(z) \eqno{(3.5)} $$ where $v$ is from
a neighbourhood of the infinity and the integration contour
encompasses this neighbourhood.

Second, a crucial simplification will be achieved by
evaluating $ {\fr T}^{(\infty )}(v)$ only at the points from
the above set $\{z_j \}$ (one has to take care that they lie
in the chosen neighbourhood). We use the formulas (2.23),
(2.21), (2.13) and (2.24) to find the variation of the tau
function $\tau \{z_j \}$, which amounts to evaluating, for a
fixed index $i,$ $$ {1\over 2\pi i} \oint dz {1\over z_i -
z} \left\{ (1 - J) {\partial \psi \{z_j\}(z)\over \partial
z} \psi ^\ast \{z_j\}(z) - J\ \psi \{z_j\}(z) {\partial \psi
 ^\ast \{z_j\}(z)\over \partial z} \right\} \equiv \partial
\left( {1\over \tau } {\cal T}(z_i)\tau \right) \eqno{(3.6)}
$$ or, after the use of (2.24),
$$ {1\over \tau } {\cal T}(z_i)\tau = {1\over 2\pi i} \oint
dz {1\over z_i - z} \left( \left( ( 1 - J) {\partial \over
\partial u} - J {\partial \over \partial z} \right) {1\over
u - z} e^{\xi (t,u) - \xi (t,z)} {\tau
(t+[z^{-1}]-[u^{-1}])\over \tau (t)} \right)^{\rm reg}_{u=z}
\eqno{(3.7)} $$
where reg implies subtracting the pole ${-1\over (u - z)^2}$
, and everything has to be reexpressed through the $\{z_j
\}$ variables. This latter task, however, will be achieved
not until the final stage of the derivation.  Now we perform
an expansion using {\it time} derivatives acting on the tau
function and thus find:  $$ \new \begin{array}{rcl} {\cal
T}(z_i) &=& {1\over 2\pi i} \oint dz { 1\over z_i - z}
\left\{ (J - {1\over 2} ) {1\over z}\sum _{r\geq 1}z^{-r-1}
{\partial \over \partial t_r} + {1\over 2}\sum
_{r,s}z^{-r-s-2} {\partial ^2\over \partial t_r\partial t_s}
\right.\\ {}&+& \left. \sum  _j n_j {1\over z_j - z}\sum
_{r\geq 1}z^{-r-1} {\partial \over \partial t_r} + {1\over
2}\sum_j {n_j + n_j^2\over (z_j - z)^2} + {1\over 2}\sum
_{^{j,k}_{j\neq k}}{n_jn_k\over (z_j - z)(z_k - z)} \right.
 \\ {}&-& \left. J\sum_j {n_j\over (z_j - z)^2} + (J -
{1\over 2}) \sum _{r\geq 1}z^{-r-2} r {\partial \over
 \partial t_r} \right\} \end{array} \eqno{(3.8)} $$
 Evaluating the residue is the crucial step which allows one
to bring (3.8) to a tractable form in terms of the $z_j$ .
As the integration contour encompasses all the points
$\{z_j\}$, the residues at both $z = z_i$ and $z = z_j$ , $j
\neq  i$, contribute to (3.8). The residue at $z_i$ consists
of the following parts: first, the terms with the
first-order pole contribute $$ \new \begin{array}{l}
\left(J - {1\over 2} - {1\over 2n_i} \right) {1\over n_i}
{1\over z_i} {\partial \over \partial z_i} - {1\over 2n_i^2
} {\partial ^2\over \partial z_i^2} +{1\over n_i}\sum
_{j\neq i}n_j {1\over z_j - z_i} {\partial \over \partial
z_i}\\ - \left( J - {1\over 2} - {1\over 2n_i}\right) \sum
_{r\geq 1}r z^{-r-2}_i {\partial  \over \partial t_r}
 -{1\over 2}\sum _{j\neq i}{n_j+n _j^2-2Jn_j\over (z_j -
z_i)^2} -{1\over 2}\sum_{{j \neq i \atop {k\neq i \atop k
\neq j}} }{n_jn_k\over (z_j - z_i)(z_k - z_i)} \end{array}
\eqno{(3.9)} $$ where we have substituted $$ \new
\begin{array}{rcl} \sum _{r,s}z^{-r-s-2}_i {\partial ^2\over
\partial t_r\partial t_s} &=& {1\over n_i^2 } {\partial
^2\over \partial z_i^2 } + {1\over n_i^2} {1\over z_i}
{\partial \over \partial z_i} - {1\over n_i} \sum _{r\geq
1}z_i^{-r-2} r {\partial \over \partial t_r}, \\ \sum
_{r\geq 1}z^{-r-1}_i {\partial \over \partial t_r} &=& -
{1\over n_i} {\partial \over \partial z_i} \end{array}
\eqno{(3.10)} $$

Next, second-order poles occur in the double sum over $j$,
$k$ in (3.8):  $$ {1\over 2\pi i} \oint dz { 1\over z_i - z}
\sum_{j\neq i} {n_j n_i \over (z_j-z)(z_i-z)}= \sum_{j\neq
i}{n_i n_j \over (z_i-z_j)^2} $$ Now, to get rid of the
$\partial /\partial t_r$-terms in (3.9) which cannot be
expressed through $\partial /\partial z_j$, we set the
coefficient in front of these equal to zero:  $$ n_i =
{1\over 2J - 1} \equiv  {1\over Q} \eqno{(3.11)} $$ Then the
contribution of the residue at $z=z_i$ equals $$ \new
\begin{array}{rcl} {\cal T}^{(i)}(z_i) = &-&{1\over 2n_i^2 }
{\partial ^2\over \partial z_i^2 } + {1\over n_i}\sum
_{j\neq i}n_j {1\over z_j - z_i} {\partial \over \partial
z_i}\\ {}&-& {1\over 2}\sum_{{j \neq i \atop {k\neq i \atop
k \neq j}} }{n_jn_k\over (z_j - z_i)(z_k - z_i)} - {1\over
2}\sum _{j\neq i}{n_j+n _j^2-2Jn_j -2n_i n_j\over (z_j -
z_i)^2} \end{array} \eqno{(3.12)} $$ Similarly,  each of the
residues at $z_j$, $j \neq  i$, contributes $$ {\cal
T}_{(j)}(z_i) = -{1\over z_j - z_i} {\partial \over \partial
z_j} + {1\over z_j - z_i} \sum _{k\neq j} {n_j n_k\over z_k
- z_j} +{1\over 2}{n_j+n_j^2-2Jn_j\over (z_i-z_j)^2}
\eqno{(3.13)} $$ and thus, $$ \new \begin{array}{rcl} {\cal
T}(z_i) &=& {\cal T}^{(i)}(z_i) + \sum_{j\neq i}{\cal
T}_{(j)}(z_i)\\ {}&=& -{1\over 2n_i^2 } {\partial ^2\over
\partial z_i^2 } + {1\over n_i} \sum_{j\neq i} {1\over z_j -
z_i} \left( n_j {\partial \over \partial z_i} - n_i
{\partial \over \partial z_j}\right) \end{array}
\eqno{(3.14)} $$ (We have used the identity $$ \sum_{j\neq
i} \sum_{{k\neq i} \atop {k\neq j}}{1\over
(z_j-z_i)}{n_jn_k\over(z_k-z_j)}= {1\over 2}\sum_{j\neq i}
\sum_{{k\neq i} \atop {k\neq j}}{n_jn_k\over
(z_j-z_i)(z_k-z_i)} ~.\quad ) $$

Now, the above treatment can be applied equally well to each
of the ${\cal T}(z_j)$, and thus (3.11) must hold for all
the $n_j$ . Finally, $$ {\cal T}(z_i) = -{Q^2\over 2}
{\partial ^2\over \partial z_i^2 } - \sum _{j\neq i} {1\over
z_j - z_i} \left( {\partial \over \partial z_j} - {\partial
\over \partial z_i}\right) \eqno{(3.15)} $$

We thus see that, indeed, in order that the ${\cal L}_{\geq
-1}-$Virasoro generators translate into the $\{z_j \}$
variables, one has to restrict the general Miwa transform
(3.1) to a Kontsevich form with all the $n_j$ equal to each
other \footnote{Restricting to only {\it integer $n_j$}
would fix two (equivalent) values $J = 0,~ 1$ of conformal
spin of the underlying abstract $ bc$ system. For our
purposes in Sect.4, however, we will need more general $n_j$
and $J$.}.

{}From the above derivation of (3.15) we see that $z_i$ is
nothing but a value taken by the spectral parameter and thus
the trick, described in {\bf 2.3}, with building up
invariants with respect to $ {\open Z}_N$ applies here as
well. That is, to perform the reduction to an $N$-KdV
hierarchy, it suffices to substitute $$ z_i \mapsto  \omega
^cz_i $$ and then sum over ${\open Z}_N$ as in (2.21)
\footnote{\ Clearly, having defined the reduced ${\cal
T}-$operator as (see (2.21)) $$ {\cal T}^{[N]}_i = {1\over
2\pi i} \oint dz {1\over z_i - z} {1\over N}\sum
^{N-1}_{c=0}\omega ^{2c} {\fr T}(\omega ^cz), $$ one
 continues this as $$ = {1\over 2\pi i} {1\over N}\sum
^{N-1}_{c=0} \oint {dz\omega ^{-c}\over z_i - \omega ^{-c}z}
\omega ^{2c}  {\fr T}(z) = {1\over 2\pi i} {1\over N}\sum
^{N-1}_{c=0}\omega ^{2c} \oint  {dz\over \omega ^cz_i - z}
{\fr T}(z) = {1\over N}\sum ^{N-1}_{c=0}\omega ^{2c} {\cal
T}(\omega ^cz_i).  $$ }. We thus arrive at $$ \new
\begin{array}{rcl} {\cal T}^{[N]}_i &\equiv&  {1\over N}\sum
^{N-1}_{c=0}\omega ^{2c} {\cal T}(z^{(c)}_i)\\
{}&=&-{Q^2\over 2} {\partial ^{_2}\over \partial z_i^2} -
\sum _{j\neq i} {1\over z^N_j - z^N_i} \left( z_j z^{N-2}_i
{\partial \over \partial z_j} - z^{N-1}_i {\partial \over
\partial z_i} \right) \end{array} \eqno{(3.16)} $$ Note that
$z^N \equiv  \zeta $ can be viewed as a spectral parameter
of the $N$-KdV hierarchy, as the $N$-KdV Lax operator $L$
(see (2.16)) satisfies $ L\psi (t,z) = z^N\psi (t,z) $.  In
terms of these variables, the operator (3.16) becomes, up to
an overall factor, $$ -{Q^2N\over 2} \zeta _i {\partial
^{_2}\over \partial \zeta_i^2 } - {Q^2(N-1)\over 2}
{\partial \over \partial \zeta _i} +\sum_{j\neq i}{1\over
\zeta_j -\zeta_i} \left(\zeta_j {\partial \over \partial
\zeta_j} - \zeta_i {\partial \over \partial \zeta_i}\right)
\eqno{(3.17)} $$ When imposing Virasoro constraints on the
$N-$reduced hierarchy, it is these $\zeta _i$ that are
candidates for eigenvalues of the ``source" matrix in a
Kontsevich-type matrix integral, at least for $Q^2 = 1$. We
consider the reformulation of the Virasoro {\it constraints}
in more detail in the next section.

\section{\it A la r\'echerche de Liouville perdu}

\leavevmode\hbox to\parindent{\hfill} Obviously now, if one
starts with the {\it Virasoro}-{\it constrained} KP
hierarchy, $i.e.,$ $$
 {\fr T}^{(\infty )}(z) = 0, \eqno{(4.1)}
$$ one ends up in the Kontsevich parametrization with the KP
Virasoro {\sl master equation} (cf. ref.\cite{[MS]}) $$
{\cal T}(z_i).\tau \{z_j \} = 0 \eqno{(4.2)} $$

The above derivation of (4.2),(3.15), with the $z_j$ (which
in the alternative approach are the eigenvalues of the
`source' matrix in a matrix integral) viewed as coordinates
on the spectral parameter complex plane, suggests an
interpretation of the master equation  in terms of a
conformal field theory living on this complex plane.  First,
it is natural to assume that (with a possible `background'
insertion at infinity) $$ \tau \{z_j \}= \lim_{ n\rightarrow
\infty} \langle \Psi (z_1) \ldots \Psi (z_n)\Phi
(\infty)\rangle \eqno{(4.3)} $$ with the pre-limit
correlators satisfying, $$ \left\{ {-Q^2\over 2} {\partial
 ^2\over \partial z_i^2 } +  \sum_{{j=1}\atop{j\neq i}}^n
{1\over z_i - z_j} \left({\partial \over \partial z_j} -
{\partial\over\partial z_i}\right) \right\} \langle \Psi
(z_1) \ldots \Psi (z_n)\Phi (\infty)\rangle = 0 \eqno{(4.4})
$$ Further, one can imagine a conformal theory of a $U(1)$
current $j(z)$ and an energy-momentum tensor $T(z)$:  $$
j(z) = \sum_{n\in {\open Z}} j_n z^{-n-1}, \qquad T(z) =
\sum_{n\in {\open Z}} L_n z^{-n-2} \eqno{(4.5)} $$ $$ \new
\begin{array}{lrcl} &\left[ j_m , j_n \right] &=& km
 \delta_{m+n,0}\\ &\left[ L_m , L_n \right] &=& (m-n)L_{m+n}
+ {d+1 \over 12}(m^3 - m)\delta_{m+n,0}\\ &\left[ L_m , j_n
\right] &=& -nj_{m+n} \end{array} \eqno{(4.6)} $$ (We have
parametrized the central charge as $d+1$). Then, in the
standard conformal field theory setting \cite{[BPZ]}, let us
look for a null vector at level 2:  $$ | \Upsilon \rangle =
\left(\alpha L_{-1}^2 + L_{-2} + \beta j_{-2} + \gamma
j_{-1}^2 + \epsilon j_{-1} L_{-1}  \right) | \Psi \rangle
\eqno{(4.7)} $$ where $\Psi$ is a primary field with
conformal dimension $\Delta$ and $U(1)$ charge $q$. We will
in fact need the specific case $\gamma = 0$. Then (4.7) is a
null vector when $$ \new \begin{array}{rclrcl} \alpha &=&
{k\over 2q^2}~,\qquad \beta &=& -{q\over k} - {1\over
2q}~,\\ \epsilon &=& -{1\over q}~,\qquad \Delta &=&
-{q^2\over k} - {1\over 2} \end{array} \eqno{(4.8)} $$ with
$q$ given by, $$ {q^2\over k} = {d-13 \pm \sqrt{(d-25)(d-1)}
\over 24} \eqno{(4.9)} $$ (so that, $$ \Delta = {1-d \mp
\sqrt{(d-25)(d-1)} \over 24} \quad .) \eqno{(4.10)} $$

Factoring the state (4.7) out from the Hilbert  space leads
in the usual manner to the equation $$ \left\{ {k\over
2q^2}{\partial^2 \over \partial  z^2 } - {1\over q} \sum_{j}
{1\over z_j - z} \left(q{\partial\over \partial z_j} -
q_j{\partial\over \partial z} \right) +{1\over
q}\sum_{j}{q\Delta_j - q_j \Delta \over (z_j - z)^2}\right\}
\langle \Psi (z) \Psi_1 (z_1) \ldots \Psi_n (z_n) \rangle =
0 \eqno{(4.11 )} $$ where $\Psi_j$ are primaries of
dimension $\Delta_j$ and $U(1)$ charge $q_j$.  In
particular, $$ \left\{ {k\over 2q^2}{\partial^2 \over
\partial  z_i^2 } + \sum_{j\neq i} {1\over z_i - z_j}
\left({\partial\over \partial z_j} - {\partial\over \partial
z_i} \right) \right\} \langle \Psi (z_1) \ldots \Psi (z_n)
\rangle = 0 \eqno{(4.12)} $$

This is to be compared with (4.4) (with the insertion at the
infinity disregarded)\footnote{ Note that the
energy-momentum tensor $T$ introduced in (4.5) appears to
have a priori nothing to do with the energy-momentum tensor
on the hierarchy we have started with. In terms of the
latter tensor, eq.(4.4) comprises the contribution of all
the positive-moded Virasoro generators, while out of $T(z)$
                       only $L_{-1}$ and $L_{-2}$ enter in
the equivalent equation (4.7).  }. We thus arrive at the
identification $$ Q^2 = -{k\over q^2} = {13-d \pm
\sqrt{(d-25)(d-1)} \over 6} \eqno{(4.13)} $$ and therefore
find ourselves in the friendly realm of minimal models
\cite{[BPZ],[DF],[FQS]}, tensored with the $U(1)$ current.
Moreover, the theory on the $z$-plane also incorporates the
gravitational dressing of the matter theory. To see this let
us first examine closer the constraint $|\Upsilon \rangle
                                   =0$.  Writing the Hilbert
space as (matter$)\otimes($current$)\equiv {\cal M}\otimes
{\cal C}$, $|\Psi \rangle =|\psi \rangle\otimes |\tilde
{\Psi}\rangle$, we introduce the matter Virasoro generators
$l_n$ by, $$ L_n = l_n + \tilde{L}_n \equiv l_n + {1\over
2k}\sum_{m\in {\open Z}}:j_{n-m}j_m :  \eqno{(4.14)} $$ They
then have central charge $d$. Singling out the
$j$-independent terms, we write $$ |\Upsilon \rangle =
\left({k\over 2q^2}l_{-1}^2 + l_{-2} \right) |\Psi \rangle +
\ldots \eqno{(4.15)} $$ By virtue of (4.13) the term written
out explicitly is by itself a null vector, and can therefore
be set to zero in ${\cal M}$. As to the other terms on the
RHS of (4.15), we substitute $$ \tilde{L}_{-1}|\tilde
{\Psi}\rangle = {q\over k}j_{-1}|\tilde {\Psi} \rangle ,
\quad \tilde{L}_{-2}|\tilde {\Psi}\rangle = \left({q\over
k}j_{-2}+{1\over 2k}j_{-1}^2\right)|\tilde {\Psi} \rangle
,\quad \tilde{L}_{-1}^2|\tilde {\Psi}\rangle = \left({q\over
k}j_{-2}+{q^2\over k^2}j_{-1}^2\right)|\tilde {\Psi} \rangle
$$ Then,
with the coefficients chosen as in (4.8), all the other
terms cancel out, and thus the ellipsis in (4.15) vanishes.

We are thus left with the null vector
$$ \left({k\over
2q^2}l_{-1}^2 + l_{-2} \right) |\psi \rangle \eqno{(4.16)}
$$ in the matter Hilbert space ${\cal M}$. The dimension of
$|\psi \rangle$ in the matter sector is found from $$ L_0
|\Psi\rangle = \left(l_0 + {1\over 2k}j_0^2 \right)
|\Psi\rangle $$ and equals $$
\delta = \Delta - {1\over 2k}q^2 =
{5-d\mp\sqrt{(1-d)(25-d)}\over 16} \eqno{(4.17)} $$ Viewing
this as the `bare' dimension we see that tensoring with the
current $j$ is equivalent to the gravitational dressing:
evaluating the gravitational scaling dimension according to
\cite{[Da],[DK]}, $$ \hat {\delta}_{\pm} =
{\pm\sqrt{1-d+24\delta}-\sqrt{1-d} \over
\sqrt{25-d}-\sqrt{1-d}} \eqno{(4.18)} $$ we find $$
\delta_+ = {3\over 8} \pm {d-4-\sqrt{(1-d)(25-d)}\over 24}
\eqno{(4.19)} $$ The sign on the RHS corresponds to that in
(4.17) and the previous formulae. In particular, choosing
the {\it lower} signs throughout, we have $$ \hat{\delta}_+
= \Delta + {1\over 2} \eqno{(4.20)} $$ Therefore, up to the
shift by $1/2$ (which seems somewhat misterious), the
dimension $\Delta$ with respect to the full Virasoro algebra
acting in the tensor product space ${\cal M}\otimes{\cal
C}$, is equal to the scaling dimension of the
gravity-dressed operators. Thus, the role of the
gravitational dressing is effectively played by tensoring
with the theory defined by $$[j_m,j_n]=km\delta_{m+n,0}, $$
$$ j_{n>0}|0\rangle = 0, \qquad j_0|0\rangle = q|0\rangle $$
with central charge 1 and negative $q^2/k$. The reason for a
current to appear at all is that it serves to represent the
hierarchy flows and thus signifies the ``hierarchical''
origin of the theory. This also provides a new insight into
the theory of completely integrable evolutions: for
Virasoro{\it -constrained} hierarchies these amount to the
Liouville dynamics in the conformal gauge.

To return to the relation with the formalism of
\cite{[Da],[DK]}, recall that the Coulomb-gas realization of
the matter theory requires introducing a scalar field
$\varphi$ with the energy-momentum tensor $$
T_{\rm m} = -{1\over 2}\partial \varphi \partial \varphi +
i{Q_{\rm m}\over 2}\partial^2\varphi \eqno{(4.21)} $$ Then
the matter central charge is equal to $1-3Q^2_{\rm m}$, and
equating this with $d$ we invert (4.13) as $$ d =
1-3{(Q^2-2)^2\over Q^2}, \eqno{(4.22)} $$ and thus $$ Q_{\rm
m}^2 = \left( Q - {2\over Q}\right)^2 \eqno{(4.23)} $$ On
the other hand, the parameter $Q$ was introduced initially
in the Virasoro constraints (2.9) (where $J={Q+1\over 2}$)
as the background charge of an abstract $bc$ system (cf.
eq.(2.13)). Now, it has to be tuned as $$ Q^2=\left({Q_{\rm
m} \pm \sqrt{Q_{\rm m}^2 + 8}\over 2}\right)^2 =
\left({Q_{\rm m}\pm Q_{\rm L} \over 2}\right)^2 =
\left({-Q_{\rm L} \mp Q_{\rm m} \over 2}\right)^2
\eqno{(4.24)} $$ where $$ Q_{\rm L} = \sqrt{Q^2_{\rm m} +8}
\eqno{(4.25)} $$ is the background charge of the `Liouville'
scalar field \cite{[DK],[Da]} with the energy-momentum
tensor $$ T_{\rm L} = -{1\over 2}\partial \phi \partial \phi
- {Q_{\rm L}\over 2}\partial^2\phi \eqno{(4.26)} $$
Equivalently, one sees that (for the respective sugns in
(4.13)), $$ Q^2=\alpha^2_{\mp} \eqno{(4.27)}
$$ where $e^{\alpha_+ \phi}$ is the gravitational dressing
of the identity operator.  This establishes the physical
meaning of $Q$ (note that $Q$ enters explicitly in the
Kontsevich transform through (3.11)). - It looks like the
$bc$ system underlying eqs.(2.13) and (2.15) describes (upon
imposing the Virasoro constraints) a `mixture' of the matter
and Liouville theories.

\section{Concluding remarks}

\leavevmode\hbox to\parindent{\hfill}{\bf 1.} It remains an
open problem to represent the $N-$reduced master equation as
a Ward identity of a matrix integral.

{\bf 2.} Our approach was based on a general construction of
Virasoro generators on the phase space of integrable
hierarchies \cite{[S4],[S6]}, and, as the $N$-KdV
hierarchies do not seem so much formally distinguished in
any way, it must apply also to other Virasoro-constrained
hierarchies, including the ``discrete" ones, $e.g$. Toda
\cite{[GMMMO],[Ma],[S4]}. This may be especially interesting
in view of the lack of a ``discretized" version of the
Kontsevich model (which does by itself seem to be
`discrete'), while, on the other hand, Virasoro constraints
on the Toda hierarchy \cite{[GMMMO],[MM]} have been shown
\cite{[S2]} to undergo a continuum limit into Virasoro
constraints on a KP hierarchy obtained from Toda also as a
result of the scaling.  It would be interesting to
investigate what kind of a Kontsevich-type matrix integral
the corresponding master equation may be derived from.

{\bf 3.} Various aspects of the conversion of Virasoro
constraints into decoupling equations would be interesting,
in particular, from the `Liouville' point of view. ?` The
Kontsevich-type matrix integral whose Ward identities
coincide with our master equation may thus provide a
candidate for a discretized definition of the Liouville
theory.

It was implicitly understood in Sect.4 that the matter
central charge $d$ should be fixed to the minimal-models
series; then factoring out the null-vector leads to a bona
fide minimal model (and our $\psi$ thus becomes the `21'
operator). Now, thinking in terms of the minimal models, how
can the {\it higher} null-vectors be arrived at starting
from the Virasoro-constrained hierarchies? If these vectors
correspond to higher symmetries of Virasoro-constrained
hierarchies, then the whole Kac table must have a relation
to the $W_{\infty}$ algebra.

\vspace{0.5cm} \noindent
{\bf {\sc Acknowledgements.}} I am grateful to O.Andreev,
A.Zabrodin and A.Mironov for useful remarks and to
A.Subbotin and R.Metsaev for valuable suggestions on the
manuscript.

              \end{document}